\documentclass[journal,12pt,onecolumn]{IEEEtran}
\usepackage{amsmath,amssymb, dsfont}
\usepackage{epsfig,latexsym,amssymb,amsmath,graphics}
\usepackage{graphicx,subfigure}
\usepackage[linesnumbered,ruled]{algorithm2e}
\usepackage{cite,color}
\usepackage{url}

\newtheorem{prop}{Proposition}

\newtheorem{cor}{Corollary}

\newtheorem{lm}{Lemma}

\newtheorem{thm}{Theorem}

\newcommand{\be}{\begin{eqnarray}}
\newcommand{\ee}{\end{eqnarray}}
\newcommand{\benn}{\begin{eqnarray*}}
\newcommand{\eenn}{\end{eqnarray*}}
\def\IR{\rm I \kern-0.20em R}
\newcommand{\utwi}[1]{\mbox{\boldmath $ #1$}}

\newcommand{\bthm}{\begin{thm}}
\newcommand{\ethm}{\end{thm}}

\newcommand{\bcor}{\begin{cor}}
\newcommand{\ecor}{\end{cor}}
\newcommand{\bprop}{\begin{prop}}
\newcommand{\eprop}{\end{prop}}
\newcommand{\blm}{\begin{lm}}
\newcommand{\elm}{\end{lm}}
\newcommand{\beq}{\begin{equation}}
\newcommand{\eeq}{\end{equation}}
\newcommand{\ber}{\begin{eqnarray}}
\newcommand{\eer}{\end{eqnarray}}

\newcommand{\bproof}{\begin{proof}}
\newcommand{\eproof}{\end{proof}}

\newcommand{\bit}{\begin{itemize}}
\newcommand{\eit}{\end{itemize}}
\newcommand{\ben}{\begin{enumerate}}
\newcommand{\een}{\end{enumerate}}
\newcommand{\bdesc}{\begin{description}}
\newcommand{\edesc}{\end{description}}
\newcommand{\beqarrn}{\begin{eqnarray*}}
\newcommand{\eeqarrn}{\end{eqnarray*}}
\newcommand{\bproofof}{\begin{proofof}}
\newcommand{\eproofof}{\end{proofof}}
\newenvironment{rem}{\begin{trivlist}\item[]{\bf
Remark:}\hspace{4mm}}{\end{trivlist}}
\newcommand{\brem}{\begin{rem}}
\newcommand{\erem}{\end{rem}}
\newenvironment{rems}{\begin{trivlist}\item[]{\bf
Remarks}\begin{itemize}}{\end{itemize}\end{trivlist}}
\newcommand{\brems}{\begin{rems}}
\newcommand{\erems}{\end{rems}}
\newtheorem{fact}{Fact}
\newcommand{\bfact}{\begin{fact}}
\newcommand{\efact}{\end{fact}}
\newtheorem{examp}{Example}
\newcommand{\bexamp}{\begin{examp}\rm}
\newcommand{\eexamp}{\end{examp}}
\newtheorem{defn}{Definition}
\newcommand{\bdefn}{\begin{defn}\rm}
\newcommand{\edefn}{\end{defn}}

\newtheorem{alg}{Algorithm}
\newcommand{\balg}{\begin{alg}}
\newcommand{\ealg}{\end{alg}}

\newtheorem{prob}{Problem}
\newcommand{\bprob}{\begin{prob}}
\newcommand{\eprob}{\end{prob}}

\newcommand{\bvtm}{\begin{verbatim}}
\newcommand{\bfig}{\begin{figure}}
\newcommand{\efig}{\end{figure}}
\newcommand{\bcen}{\begin{center}}
\newcommand{\ecen}{\end{center}}

\long\def\comment#1{}

\def \n2{{N_0 \over 2}}

\def \h5{\hspace{0.5in}}

\newcommand{\bb}{{\utwi{b}}}

\newcommand{\bk}{{\utwi{k}}}

\newcommand{\bv}{{\utwi{v}}}

\newcommand{\bz}{{\utwi{z}}}

\newcommand{\bN}{{\utwi{N}}}

\newcommand{\bY}{{\utwi{Y}}}
\newcommand{\bZ}{{\utwi{Z}}}

\newcommand{\dff}{\stackrel{\triangle}{=}}

\renewcommand{\baselinestretch}{1}

\def\IR{\mathbb R}

\renewcommand{\baselinestretch}{1.6}

\newtheorem{theorem}{Theorem}

\title{Signal Detection under Short-Interval Sampling of Continuous Waveforms for Optical Wireless Scattering Communication}

\author{Difan Zou, Chen Gong and Zhengyuan Xu
\thanks{This work was supported in part by National 973 Program of China (Grant No. 2013CB329201), Shenzhen Peacock Plan (No.108170036003286) and the Fundamental Research Funds for the Central Universities.}
\thanks{The authors are with Key Laboratory of Wireless-Optical Communications, Chinese Academy of Sciences, School of Information Science and Technology, University of Science and Technology of China, Hefei, Anhui 230027, China. Z. Xu is also with Shenzhen Graduate School, Tsinghua University, Shenzhen, China. Email:  knowzou@mail.ustc.edu.cn,\{cgong821,xuzy\}@ustc.edu.cn.}}
\date{}

\begin{document}

\maketitle{}

\renewcommand{\baselinestretch}{1.3}

\begin{abstract}
In optical wireless scattering communication, received signal in each symbol interval is captured by a photomultiplier tube (PMT) and then sampled through very short but finite interval sampling. The resulting samples form a signal vector for symbol detection. The upper and lower bounds on transmission rate of such a processing system are studied. It is shown that the gap between two bounds approaches zero as the thermal noise and shot noise variances approach zero. The maximum a posteriori (MAP) signal detection is performed and a low computational complexity receiver is derived under piecewise polynomial approximation. Meanwhile, the threshold based signal detection is also studied, where two threshold selection rules are proposed based on the detection error probability and the Kullback-Leibler (KL) distance. For the latter, it is shown that the KL distance is not sensitive to the threshold selection for small shot and thermal noise variances, and thus the threshold can be selected among a wide range without significant loss from the optimal KL distance. The performances of the transmission rate bounds, the signal detection, and the threshold selection approaches are evaluated by the numerical results.

\end{abstract}

\renewcommand{\baselinestretch}{1.6}

\thispagestyle{empty}

\section{Introduction}
Due to the potential large bandwidth and no electromagnetic radiation, optical wireless communication shows great promise for the future wireless communications\cite{gagliardi1976optical}. It can be deployed for the applications where the radio-frequency (RF) radiation is prohibited, for example in a hospital or aircraft cabin where the electromagnetic radiation is of particular concern. In some outdoor scenarios,, the line-of-sight (LOS) link between the transmitter and the receiver may be blocked by an obstacle, or cannot be guaranteed due to the application requirements. The solution is to utilize the non-line-of-sight (NLOS) optical scattering communication \cite{xu2008ultraviolet,ding2009modeling}, typically in the ultra-violet spectrum. For the NLOS communication, the transmitting and receiving directions are not required to be perfectly aligned, which expands the application range beyond the LOS links. The NLOS optical scattering communication channels and systems have been extensively studied from both experimental perspective \cite{ding2009modeling,ding2010path,xiao2011non,liao2015uv}, and semi-analytical perspective \cite{wang2010non,he2010performance,he2010achievable}.

Due to the large path loss between the transmitter and the receiver, the received signals are characterized by discrete photoelectrons. Using a photon-counting receiver, the number of detected photoelectrons satisfies a Poisson distribution, which forms a Poisson channel. The capacity of the Poisson channel has been studied in \cite{wyner1988capacity,frey1991information,lapidoth2009capacity}. Recently, the capacity and the optimized source distribution for the discrete-time Poisson (DTP) channel have been investigated in \cite{cao2014capacity1}, \cite{cao2014capacity2}. The capacity and signal detection of the free-space MIMO optical wireless communication system based on the ideal photon-counting receiver and maximum-likelihood (ML) detection have been investigated in \cite{wilson2005optical}, \cite{wilson2005free}. The base-band digital signal processing and the coded modulation have been studied in \cite{georghiades1994synchronizable,chatzidiamantis2010generalized,nguyen2010coded}. Besides, the communication channel modeling and signal processing has been investigated, in the aspects of the inter-symbol interference modeling \cite{ScatteringISI}, the relay protocol for the photon-counting receiver \cite{he2010non}, \cite{gong2015non}, and the linear receiver for the SIMO scattering communication with Poisson shot noise \cite{ScatteringSIMO}.

Note that the capacity for the Poisson channel is achieved as interval for counting the detected photoelectrons becomes arbitrarily small. However, this cannot be realized in a practical receiver. On the other hand, an ideal discrete photon-counting receiver is difficult to realize, while typically a continuous waveform processing (WP) receiver consisting of a photomultiplier tube (PMT) and postprocessing circuits is employed. The WP receiver amplifies each detected photoelectron to a series of electrons, with the additive signal dependent shot noise and signal independent thermal noise. It becomes interesting to investigate the achievable transmission rate and the signal detection performance of the WP receiver, where interval for measuring the PMT outputs cannot be arbitrarily small.

In this work, we study the transmission rate and the signal detection for such type of practical WP receiver from three perspectives, the transmission rate, the signal detection, and the non-ideal photon-counting receiver based on the hard-decision with a preset threshold. Assume a small but finite time processing interval for measuring the PMT outputs, such that PMT output signals within each symbol duration form a vector consisting of the output signals within the intervals. To the best of our knowledge, such a model has not been analyzed before. We investigate the upper and lower bounds on the transmission rate of such type of channel. We also study the maximum a posteriori (MAP) signal detection for the transmitted symbol, and approximation at reduced computational complexity.  Moreover, we consider a non-ideal photon-counting receiver, where a preset threshold is employed for the hard decision on whether a photoelectron exists in each interval. The optimal threshold selection rule is investigated based on two criteria, to minimize the detection error probability and to maximize the Kullback-Leibler (KL) distance. We further prove that for small shot and thermal noise variances, the detection threshold can be selected among a wide range without significant loss from the optimal KL distance. The performances of the proposed transmission rate bounds, the signal detection, and the optimal threshold selection criteria for the non-ideal photon-counting receiver are evaluated by the numerical results.

The remainder of this paper is organized as follows. In Section II, we present the model of the NLOS optical wireless scattering communication system and the PMT output signal. In Section III, we investigate the upper and lower bounds on the communication rate. In Section IV, we consider the MAP detection of the transmitted symbol, and propose a reduced computational complexity receiver. In Section V, we provide the signal model for the non-ideal photon-counting receiver, and obtain the optimal threshold based on the detection error probability and the KL distance. Numerical results are given in Section VI. Finally, Section VII provides the concluding remarks.

\section{Channel Model}
\subsection{Optical Wireless Scattering Communication with WP receiver}
Consider an optical wireless scattering communication system, typically in the ultra-violet (UV) spectrum. Due to the large path loss of the scattering by the particles and aerosol in the atmosphere, the intensity of received signal becomes extremely weak, where the signals are characterized by the discrete photoelectrons. The detected photoelectrons consist of two components, the desired signal component and the background radiation component.  The numbers of photoelectrons for both components satisfy Poisson distributions.

Assume that the transmitter adopts on-off keying (OOK) modulation, where $X \in \{0,1\}$ denotes the transmitted symbol. Let $P$ denote the transmission power for $X = 1$ when the transmission is on, and $g$ denote the link gain between the transmitter and the receiver. Let $h$ and $\nu$ denote the Planck's constant and the frequency of the optical carrier, respectively, such that the energy per photon is given by $h\nu$. Let $\eta$ denote the detector quantum efficiency as the ratio between the number of photoelectrons over the number of received photons; and $\tau$ denote the length of the OOK symbol duration. The mean number of detected photoelectrons for the desired signal component corresponding to the transmission power $P$, denoted as $\lambda_s$, is given by
\begin{eqnarray} \label{equ.SgnComponent}
\lambda_s = \frac{Pg\eta \tau}{h\nu}.
\end{eqnarray}

Let $\lambda_b$ denote mean number of detected photoelectrons for the background radiation in each symbol duration. The number of detected photoelectrons, denoted as $N$, satisfies the following Poisson distribution,
\begin{eqnarray} \label{equ.PoissonDistribution}
\mathbb{P}(N = n|X=1) = \frac{(\lambda_s + \lambda_b)^n}{n!}e^{-\lambda_s - \lambda_b}.
\end{eqnarray}

Consider a practical PMT detector that amplifies each detected photoelectron into a series of electrons. Let $A$ and $e$ denote the amplification factor and the charge of each electron, respectively. The PMT output signal upon detecting $n$ photoelectrons, denoted as $z$, is given by
\begin{equation}
z=nAe+v,
\end{equation}
where $v$ denotes the additive Gaussian noise including the shot noise and the thermal noise. Let $\sigma^2$ and $\sigma^2_0$ be the variances of the shot noise per photoelectron and the thermal noise, respectively. Note that the additive noise $v$ satisfies the Gaussian distribution ${\cal N}(0, n\sigma^2 + \sigma^2_0)$ with mean zero and variance $ n\sigma^2 + \sigma^2_0$.  We have the following on the shot noise per photoelectron the thermal noise variances,
\begin{eqnarray}
\sigma^2=\xi^2A^2e^2,\quad \sigma_0^2=\frac{2k_eT^o\tau}{R},
\end{eqnarray}
where $\xi$ denotes the PMT spreading factor; $k_e$ denotes the Boltzmann constant; $T_o$ denotes the temperature ($K$); and $R$ denotes the load resistance. Let $G(z;\mu, \kappa^2)$ denote the Gaussian probability distribution function (PDF) with mean $\mu$ and variance $\kappa^2$. The pdf of $z$ upon detecting $n$ photoelectrons, denoted as $p(z|N=n)$, is given as follows,
\begin{eqnarray}\label{eq.output_gauss}
p(z|N=n)&=&G(z;nAe,n\sigma^2+\sigma_0^2) \nonumber \\
&=&\frac{1}{\sqrt{2\pi\left(n\sigma^2+\sigma_0^2\right)}}e^{-\frac{\left(z-nAe\right)^2}{2\left(n\sigma^2+\sigma_0^2\right)}}\dff G_n(z).
\end{eqnarray}
 The pdf of the PMT output signal, denoted as $p(z|\lambda_s + \lambda_b)$, is given by
\begin{eqnarray} \label{equ.OutputPMT2}
p(z|\lambda_s + \lambda_b) &=& \sum^{+\infty}_{n=0}\mathbb{P}(N=n)p(z|N=n) \nonumber \\
&=& \sum^{+\infty}_{n=0}\frac{(\lambda_s + \lambda_b)^n}{n!}e^{-\lambda_s - \lambda_b}G(z; nAe, n\sigma^2 + \sigma^2_0).
\end{eqnarray}

\subsection{The WP receiver with a Finite Processing Time Interval}
Note that for the Poisson channel, the capacity is achieved when the receiver can count the number of detected photoelectrons in an arbitrarily small interval. However, this cannot be realized by a practical receiver, where the interval for counting the detected photoelectrons cannot be arbitrarily small. In this work, we assume that the minimum processing interval for the PMT output is given by $\frac{\tau}{M}$ for some integer $M$. This can be implemented via adopting a high-rate sampling to the PMT output signals, and detecting the transmitted symbol based on the sampled signals within each minimum processing interval.

For each symbol duration, let $\bz \dff [z_1, z_2, \dots, z_M]$ denote the output analog signal vector of the WP receiver in the $M$ intervals. We need to decode the information symbol $X$ based on the signal vector $\bz$ of the $M$ intervals. Note that the number of detected photoelectrons for each interval satisfies a Poisson distribution, with means $\gamma_t \dff \frac{\lambda_s + \lambda_b}{M}$ and $\gamma_b \dff \frac{\lambda_b}{M}$ for OOK symbols $X = 1$ and $X = 0$, respectively. The conditional pdfs for $z_m$ given $X = 0$ and $X = 1$, denoted as $p(z_m|X = 1)$ and $p(z_m|X = 0)$, respectively, are given by
\begin{eqnarray}\label{eq.likelihood_twocase}
p(z_m|X=0)&=&\sum_{n=0}^{+\infty}\frac{\gamma_b^n}{n!}e^{-\gamma_b}G\left(z;nAe,n\sigma^2+\frac{\sigma^2_0}{M}\right), \nonumber \\
p(z_m|X=1)&=&\sum_{n=0}^{+\infty}\frac{\gamma_t^n}{n!}e^{-\gamma_t}G\left(z;nAe,n\sigma^2+\frac{\sigma^2_0}{M}\right).
\end{eqnarray}

Assume that given $X$, the output signals $z_1$, $z_2$, ..., $z_M$ are independent of each other. This can be justified by the independent Poisson arrival events and additive Gaussian noise in different intervals. Thus the pdfs of the output signal $\bz$ for $X = 0$ and $X = 1$, denoted as $p(\bz|X = 0)$ and $p(\bz|X = 1)$, respectively, are given by
\begin{eqnarray} \label{equ.OutputPMTVec}
p(\bz|X=0) &=& \prod^M_{m=1} p(z_m|X=0), \nonumber \\
p(\bz|X=1) &=& \prod^M_{m=1} p(z_m|X=1),
\end{eqnarray}
where the conditional pdfs $ p(z_m|X=0)$ and $p(z_m|X=1)$ for $1 \leq m \leq M$ are given by (\ref{eq.likelihood_twocase}).
\subsection{The Single-Photon Approximation}
For fixed symbol duration such that $\lambda_s$ and $\lambda_b$ are all fixed, we consider sufficiently small $\frac{\tau}{M}$ such that both $\gamma_t$ and $\gamma_b$ are small. In such a scenario, for Poisson distributions with means $\gamma_t$ and $\gamma_b$, the probability for detecting more than one photoelectrons becomes negligible. Then the Poisson distributions can be approximated by Bernoulli distributions. The probability for the detected photoelectron number $N_m$ in a length-$\frac{\tau}{M}$ interval is given by,
\begin{eqnarray}
\mathbb{P}(N_m=0|X=0)&=&e^{-\gamma_b}\approx1-\gamma_b = 1 - \mathbb{P}(N_m = 1|X=0), \nonumber \\
\mathbb{P}(N_m=0|X=1)&=&e^{-\gamma_t}\approx1-\gamma_t = 1 - \mathbb{P}(N_m = 1|X=1).
\end{eqnarray}
Based on the above approximation and $G_n(x)$ defined in (\ref{eq.output_gauss}), for $X = 0$ and $X = 1$, the pdfs of the output signal $z_m$, $1 \leq m \leq M$, are given by
\begin{eqnarray}\label{eq.likelihood0}
p\left(z_m|X=0\right)&\approx&(1-\gamma_b)G\left(z_m;0,\sigma_0^2\right)+\gamma_b G\left(z_m;Ae,\sigma^2+\sigma_0^2\right) \nonumber \\
&=&(1-\gamma_b)G_0(z_m)+\gamma_b G_1(z_m), \nonumber \\
p\left(z_m|X=1\right)&\approx&(1-\gamma_t)G_0(z_m)+\gamma_t G_1(z_m).
\end{eqnarray}

The single-photon approximation can be justified by the observation on the PMT output signal from the oscilloscope, which is characterized by the discrete pulses corresponding to the detected photoelectrons. Typically if dividing the entire symbol duration into $M$ small intervals for a large $M$, there is at most one pulse in each length-$\frac{\tau}{M}$ interval. The motivation of this work is to detect the OOK symbol $X$ based on the output analog signal in each interval. We address such a problem in three perspectives, the communication rate for the OOK modulation, the signal detection, and the non-ideal photon-counting receiver based on the output signal vector $\bz$.

\section{The Transmission Rate}
Let $Z_1,Z_2,\dots,Z_M$ denote the stochastic variable versions of the signals $z_1$, $z_2$, ..., $z_M$. In this Section, we first investigate the mutual information $I(X;Z_m)$ for each interval. It also represents the mutual information per OOK symbol for sufficiently small symbol duration, such that $\lambda_s$ and $\lambda_b$ are sufficiently small for $M = 1$. We provide the upper and lower bounds on the mutual information $I(X;Z_m)$. Finally we extend the upper and lower bounds to the mutual information $I_M$ for the $M$ intervals.

\subsection{Upper and Lower Bounds on the Transmission Rate for Single Interval}
Consider the priori probability,  $\mathbb{P}(X=1)=w$, and $\mathbb{P}(X=0)=1-w$. In this subsection, without loss of generality we set $M = 1$. Let $p_0(z) \dff p(z|X=0)$, $p_1(z) \dff p(z|X=1)$; and $p(z) \dff (1-w)p_0(z) + wp_1(z)$ denote the pdf of $z$. Based on the single-photon approximation, the communication rate for a single interval is given by
\begin{eqnarray}
I(X;Z_1)&=&H(Z_1)-H(Z_1|X)\nonumber \\
&=&-\int_{-\infty}^{\infty}p(z)\log p(z)dz + (1-w)\int_{-\infty}^{\infty}p_0(z)\log p_0(z)dz\nonumber\\
&&\ \ \ \ +w\int_{-\infty}^{\infty}p_1(z)\log p_1(z)dz.
\label{capacity}
\end{eqnarray}

Consider the transmitted OOK symbol $X$, the number of detected photoelectrons $N_1$, the PMT output analog signal $Z_1$, and the number of detected photoelectrons inferred from the analog signal $Z_1$, denoted as $\hat N_1$. We have that $X \rightarrow N_1  \rightarrow Z_1  \rightarrow \hat N_1$ forms a Markov chain. Based on this, the mutual information $I(X;Z_1)$ can be bounded as follows:
\begin{equation}\label{eq.bounds}
I(X;\hat N_1)\leq I(X;Z_1)\leq I(X;N_1).
\end{equation}
In the following we analyze the gap between the upper bound and the lower bound. The main result is that the gap attenuates  with the summation of shot noise variance and the thermal noise variance in a super-power manner.

Note that both transitions $X \rightarrow N_1$ and $X \rightarrow\hat N_1$ form binary asymmetric channels. Such type of channel can be characterized by the conditional probabilities $p_{01}$ and $p_{11}$ for the output symbol $1$ given OOK symbols $X = 0$ and $X = 1$. The mutual information is given by
\begin{eqnarray}\label{bac_capacity}
C_1\left(p_{01},p_{11},w\right)&\dff&H(Y)-H(Y|X) \nonumber \\ &=&H_2\left(wp_{11}+(1-w)p_{01}\right)-wH_2\left(p_{11}\right)-(1-w)H_2\left(p_{01}\right),
\end{eqnarray}
where the entropy function $H_2(x)=-x\log x-(1-x)\log(1-x)$. Note that the bounds $I(X;N_1)$ and $I(X;\hat N_1)$ can be obtained based on setting the conditional probabilities $p_{01}$ and $p_{11}$ to the corresponding values.

For the binary channel $X \rightarrow N_1$, based on the analysis in Section II.C, we have the following,
\begin{eqnarray}\label{count_error}
p_{11} = \mathbb{P}(N_1=1|X=1)=\gamma_t, \nonumber \\
p_{01} = \mathbb{P}(N_1=1|X=0)=\gamma_b.
\end{eqnarray}

For the binary channel $X \rightarrow \hat N_1$, the detection of $\hat N_1$ from the PMT output signal $Z_1$ could be performed according to the MAP criterion. However, the complicated expression of the detection threshold is not tractable for further analysis. Thus we resort to a more tractable detection of $\hat N_1$ from the output signal $Z_1$ based on a simple detection threshold $z_{th}$, such that $\hat N_1 = 1$ is detected if $z_1 > z_{th}$, and $\hat N_1 = 0$ is detected otherwise. The simple threshold can still lead to the super-linear attenuation of the gap between the upper bound and the lower bound. The typical range of the detection threshold $z_{th}$ is given by $0 < z_{th} < Ae$, for example $z_{th} = \frac{Ae}{2}$.

In this scenario, the corresponding conditional probabilities $p_{01}$ and $p_{11}$ are given by
\begin{eqnarray}\label{detect_error}
p_{11} &=& t_1 \dff \int_{z_{th}}^{\infty}f(z|x=1)dz \nonumber \\
&=&\left(1-\gamma_t\right)Q\left(\sqrt{M}\frac{z_{th}}{\sigma_0}\right)+\gamma_tQ\left(\frac{z_{th}-Ae}{\sqrt{\frac{\sigma^2_0}{M}+\sigma^2}}\right) ,\nonumber \\
p_{01} &=& t_0 \dff \int_{z_{th}}^{\infty}f(z|x=0)dz \nonumber \\
&=&\left(1-\gamma_b\right)Q\left(\sqrt{M}\frac{z_{th}}{\sigma_0}\right)+\gamma_bQ\left(\frac{z_{th}-Ae}{\sqrt{\frac{\sigma^2_0}{M}+\sigma^2}}\right), \end{eqnarray}
where $Q(\cdot)$ denotes the following Gaussian-$Q$ function
\begin{equation}
Q(x)=\int_x^{+\infty}\frac{1}{\sqrt{2\pi}}e^{-\frac{u^2}{2}} d u.
\end{equation}

It is easily seen that conditional probabilities $t_0$ and $t_1$ approach $\gamma_b$ and $\gamma_t$ as the shot noise variance $\sigma^2$ and the thermal noise variance $\sigma^2_0$ approach zero. The upper and lower bounds on the communication rate $I(X;Z_1)$ [c.f. (\ref{capacity})] can be given in the following result.

\newtheorem{proposition}{{Proposition}}
\begin{proposition}
The upper and lower bounds on the transmission rate $I(X;Z_1)$ are given as follows,
\begin{eqnarray}
C_1(t_0,t_1,w) \leq I(X;Z_1) \leq C_1(\gamma_b,\gamma_t,w).
\end{eqnarray}
Moreover, the upper and lower bounds are asymptotically tight as the shot noise variance and the thermal noise variance approach zero, i.e.,
\begin{eqnarray}
\lim_{\sigma^2, \sigma^2_0 \rightarrow 0} C_1(\gamma_b, \gamma_t, w) - C_1(t_0, t_1, w) = 0.
\end{eqnarray}
\end{proposition}
$\hfill \Box$

An upper bound on the gap $C(\gamma_b, \gamma_t, w) - C(t_0, t_1, w)$ can be obtained based on the following lemma.

\newtheorem{lemma}{Lemma}
\begin{lemma}
For the Markov chain $X \rightarrow N_1 \rightarrow Z_1 \rightarrow \hat N_1$, we have that
\begin{eqnarray}
I(X;N_1) - I(X;\hat N_1) \leq H(N_1| \hat N_1).
\end{eqnarray}
\begin{proof}
We have the following on the mutual information gap $I(X;N_1) - I(X;\hat N_1)$,
\begin{eqnarray}
I(X;N_1) - I(X;\hat N_1)&=&H(X)- H(X| N_1)-\left(H(X)-H(X|\hat N_1)\right) \nonumber \\
&=&H(X|\hat N_1)-H(X|N_1) \nonumber \\
&=&H(X|\hat N_1)-H(X|N_1,\hat N_1) \nonumber \\
&=&I(X;N_1|\hat N_1) \nonumber \\
&\leq& H(N_1 | \hat N_1).
\end{eqnarray}
\end{proof}
\end{lemma}

In the following we analyze the convergence of the conditional entropy $H(N_1|\hat N_1)$. Letting $x_1 \dff \frac{z_{th} - Ae}{\sqrt{\sigma^2+\frac{\sigma^2_0}{M}}}$ and $x_2 \dff \frac{z_{th}}{\sigma_0}$, we have that
\begin{eqnarray}
\mathbb{P}(\hat N_1 = 1|N_1 = 1) &=& Q(x_1), \nonumber \\
\mathbb{P}(\hat N_1 = 1|N_1 = 0) &=& Q(x_2).
\end{eqnarray}
Letting $r_0 \dff \mathbb{P}(N_1 = 0)$ and $r_1 \dff \mathbb{P}(N_1 = 1)$, we have the following on the conditional probabilities $\mathbb{P}(N_1 = 1|\hat N_1)$ for $\hat N_1 = 0$ and $\hat N_1 = 1$,
\begin{eqnarray}\label{eq.prob1}
\mathbb{P}(N_1=1|\hat N_1=0)&=&\frac{\Big(1 - Q(x_1)\Big)r_1}{\Big(1 - Q(x_1)\Big)r_1 + \Big(1 - Q(x_2)\Big)r_0} , \nonumber \\
\mathbb{P}(N_1=1|\hat N_1=1)&=&\frac{Q(x_1)r_1}{Q(x_1)r_1 + Q(x_2)r_0} .
\end{eqnarray}

It can be proved that the conditional entropy $H(N_1|\hat N_1)$ attenuates with the variances $\sigma^2$ and $\sigma^2_0$
in the super-power order for both $\hat N_1 = 0$ and $\hat N_1 = 1$. More specifically, we have the following result.

\begin{theorem}
For any $\mu > 0$, we have the following on the asymptotical results on the conditional entropy $H(N_1|\hat N_1)$,
\begin{eqnarray}
\lim_{\sigma^2, \sigma^2_0 \rightarrow 0} \frac{H(N_1|\hat N_1 = 0)}{(\sigma^2 + \frac{\sigma^2_0}{M})^\mu} &=& 0, \label{equ.AsympH0} \\
\lim_{\sigma^2, \sigma^2_0 \rightarrow 0} \frac{H(N_1|\hat N_1 = 1)}{(\sigma^2 + \frac{\sigma^2_0}{M})^\mu} &=& 0. \label{equ.AsympH1}
\end{eqnarray}
\begin{proof}
We provide the proof on $H(N_1|\hat N_1 = 0)$ as follows.  Note that we have the following,
\begin{eqnarray}
\hspace{-15mm}H(N_1|\hat N_1 = 0) &=& -\frac{\Big(1 - Q(x_1)\Big)r_1}{\Big(1 - Q(x_1)\Big)r_1 + \Big(1 - Q(x_2)\Big)r_0}\log_2\frac{\Big(1 - Q(x_1)\Big)r_1}{\Big(1 - Q(x_1)\Big)r_1 + \Big(1 - Q(x_2)\Big)r_0} \nonumber \\
\hspace{-15mm}&& \ \ \ -\frac{\Big(1 - Q(x_2)\Big)r_0}{\Big(1 - Q(x_1)\Big)r_1 + \Big(1 - Q(x_2)\Big)r_0}\log_2\frac{\Big(1 - Q(x_2)\Big)r_0}{\Big(1 - Q(x_1)\Big)r_1 + \Big(1 - Q(x_2)\Big)r_0}.
\end{eqnarray}

As $\frac{\ln (1+x)}{1+x}<\frac{\ln x}{x}$ and $\ln(1+x)<x$ for $x>0$, we have following
\be
&&-\frac{\Big(1 - Q(x_1)\Big)r_1}{\Big(1 - Q(x_1)\Big)r_1 + \Big(1 - Q(x_2)\Big)r_0}\log_2\frac{\Big(1 - Q(x_1)\Big)r_1}{\Big(1 - Q(x_1)\Big)r_1 + \Big(1 - Q(x_2)\Big)r_0} \nonumber \\
&&\ \ \  <-\frac{1}{\ln 2}\frac{\Big(1 - Q(x_1)\Big)r_1}{\Big(1 - Q(x_2)\Big)r_0}\ln\frac{\Big(1 - Q(x_1)\Big)r_1}{\Big(1 - Q(x_2)\Big)r_0},
\ee
\begin{eqnarray}
&&-\frac{\Big(1 - Q(x_2)\Big)r_0}{\Big(1 - Q(x_1)\Big)r_1 + \Big(1 - Q(x_2)\Big)r_0}\log_2\frac{\Big(1 - Q(x_2)\Big)r_0}{\Big(1 - Q(x_1)\Big)r_1 + \Big(1 - Q(x_2)\Big)r_0} \nonumber \\
&&\ \ \  < \frac{1}{\ln 2}\frac{r_0r_1\Big(1 - Q(x_2)\Big)\Big(1 - Q(x_1)\Big)}{\left[\Big(1 - Q(x_1)\Big)r_1 + \Big(1 - Q(x_2)\Big)r_0\right]^2}.
\end{eqnarray}
Note that $x_1 \rightarrow -\infty$ and $x_2 \rightarrow +\infty$ as $\sigma^2 + \sigma^2_0$ approaches zero. Thus, we have that $1 - Q(x_2) \rightarrow 1$ and the following bounds for $x_1 < 0$,
\begin{eqnarray}
-\frac{1}{x_1}e^{-\frac{x^2_1}{2}} < 1 - Q(x_1) < -\frac{x_1}{1 + x^2_1}e^{-\frac{x^2_1}{2}}.
\end{eqnarray}
Thus, the limit of condition entropy $H(N_1|\hat N_1 = 0)$ is bounded by
\be
\lim_{x_2 \rightarrow +\infty}H(N_1|\hat N_1 = 0)&<&\frac{r_1}{r_0\ln2}\left[-\Big(1 - Q(x_1)\Big)\ln\frac{\Big(1 - Q(x_1)\Big)r_1}{r_0}+1-Q(x_1)\right] \nonumber \\
&<&\frac{r_1}{r_0\ln2}\frac{x_1e^{-\frac{x^2_1}{2}}}{1 + x^2_1}\left(1+\frac{x_1^2}{2}+\ln x_1-\ln\frac{r_1}{r_0}\right)
\ee
For $x_1\rightarrow -\infty$, note that $e^{-\frac{x^2_1}{2}}$ contains the exponential term $-\frac{(z_{th} - Ae)^2}{2(\sigma^2 + \frac{\sigma^2_0}{M})}$ which attenuates with the variances in the super-power order. Based on the above analysis, the rest proof of (\ref{equ.AsympH0}) can be easily derived using the standard arguments on the mathematical analysis.

The proof of (\ref{equ.AsympH1}) for $H(N_1|\hat N_1 = 1)$ are similar to that for $H(N_1|\hat N_1 = 0)$, and thus omitted here.
\end{proof}
\end{theorem}

\subsection{Upper and Lower Bounds on the Transmission Rate for Multiple Intervals}
We extend the previous results to a vector case where M intervals yield a vector output. Let $\bz \dff\left[z_1,z_2,...,z_M\right]$ denote the analog PMT output in the $M$ intervals. Let $\bZ$ denote the stochastic representation of $\bz$ for the mutual information, such that
\begin{equation}\label{eq.capacity_Z}
I(X;\bZ)=H(\bZ)-H(\bZ|X).
\end{equation}
Similar to that for single interval, let $\bN$ denote the vector of the number of detected photoelectrons in the $M$ intervals, and $\hat\bN$ denote the vector of photoelectrons inferred from the signal vector $\bZ$. It can be proved that $X \rightarrow \bN \rightarrow \bZ \rightarrow \hat\bN$ forms a Markov chain. The mutual information $I(X;\bZ)$ can be bounded as follows,
\begin{eqnarray}
I(X;\hat \bN)\leq I(X;\bZ)\leq I(X;\bN).
\end{eqnarray}

Note that given $X$, the components of $\bN$ and $\hat \bN$ are independent and identically distributed satisfying Bernoulli distributions. The mutual information of such type of channel can be characterized by the conditional probabilities $p_{01}$ and $p_{11}$ for each component being one given the OOK symbol $X = 0$ and $X = 1$, respectively. Let $\bY$ be the output of such a channel. The conditional probabilities $p(\bY|X = 0)$ and $p(\bY|X = 1)$ are given by
\begin{eqnarray}
p\left(\bY|X=0\right)=(1 - p_{01})^{M - w(\bY)}p_{01}^{w(\bY)}, \\
p\left(\bY|X=1\right)=(1 - p_{11})^{M - w(\bY)}p_{11}^{w(\bY)},
\end{eqnarray}
where $w(\bY)$ denotes the weight of $\bY$. The probability $p(\bY)=(1-w)p(\bY|X=0)+wp(\bY|X=1)$,
and the entropy $\bY$ is given by
\begin{eqnarray}
H(\bY)&=&-\sum_{\bY\in\left\{0,1\right\}^M}p(\bY)\log p(\bY) ,
\end{eqnarray}
The conditional entropy $H(\bY|X)$ is given by
\begin{eqnarray}\label{eq.conditional_vector_entropy}
H(\bY|X)&=&-\mathbb{P}(X=0)\sum_{\bY\in\left\{0,1\right\}^M}p(\bY|X=0)\log p(\bY|X=0) \nonumber \\
&& \ \ \ \ \mathbb{P}(X=1)\sum_{\bY\in\left\{0,1\right\}^M}p(\bY|X=1)\log p(\bY|X=1) \nonumber \\
&=&M\left[wH_2\left(p_{11}\right)+(1-w)H_2\left(p_{01}\right)\right].
\end{eqnarray}
The mutual information for such binary asymmetric channel is given by $C_M\left(p_{01},p_{11},w\right)\dff H(\bY)-H(\bY|X)$.  The upper and lower bounds can be obtained via setting $[p_{01}\ \ p_{11}] = [\gamma_b\ \ \gamma_t]$
and $[p_{01}\ \ p_{11}] = [t_0\ \ t_1]$ for $I(X;\bN)$ and $I(X;\hat\bN)$, respectively. The parameters $[t_0\ \ t_1]$ have been given in Section III.A.

Based on the above analysis, we have the following result on the upper and lower bounds on the mutual information $I(X;\bZ)$.

\begin{theorem}
The upper and lower bounds on the transmission rate $I(X;\bZ)$ are given as follows,
\begin{eqnarray}
C_M(t_0,t_1,w) \leq I(X;\bZ) \leq C_M(\gamma_b,\gamma_t,w).
\end{eqnarray}
Moreover, the upper and lower bounds are asymptotically tight as the shot and thermal noise variances approach zero, i.e.,
\begin{eqnarray}
\lim_{\sigma^2, \sigma^2_0 \rightarrow 0} C_M(\gamma_b,\gamma_t, w) - C_M(t_0, t_1, w) = 0.
\end{eqnarray}
\end{theorem}

$\hfill \Box$

Similar to Lemma $1$ for the single interval scenario, we have following result,
\begin{eqnarray}
I(X;\bN) - I(X;\hat \bN) \leq H(\bN|\hat\bN).
\end{eqnarray}

Similar to the scenario for single interval, we have the following results on the conditional entropy $H(\bN|\hat\bN)$.
The proof follows the same procedure as that of Theorem~$2$, and thus omitted here.

\begin{theorem}
For any $\mu > 0$, we have the following asymptotical result on the conditional entropy $H(\bN|\hat\bN)$,
\begin{eqnarray}
\lim_{\sigma^2, \sigma^2_0 \rightarrow 0} \frac{H(\bN|\hat \bN)}{(\sigma^2 + \sigma^2_0)^\mu} = 0. \label{equ.AsympHH0}
\end{eqnarray}
\end{theorem}
$\hfill \Box$

\section{The detection}
In this Section, we investigate the MAP detection of the OOK symbols $X$ based on the PMT output signal $\bz$. We also provide a piecewise polynomial approximation for the log-likelihood to reduce the computational complexity.
\subsection{The MAP Detection Framework}
For the MAP detection of $X$ based on the PMT output signal in the $M$ intervals $\bz$, symbol $X = 1$ is detected if the following is satisfied
\begin{eqnarray}\label{equ.DetectionMAP01}
\log\frac{\mathbb{P}(X = 1|\bz)}{ \mathbb{P}(X = 0|\bz)} > 0;
\end{eqnarray}
and symbol $X = 0$ is detected otherwise. Such a detection rule is equivalent to the rule that $X = 1$ is detected if and only if the following is satisfied
\begin{eqnarray}\label{equ.DetectionMAP02}
\log \frac{\mathbb{P}(\bz |X = 1)}{ \mathbb{P}(\bz|X = 0)} > \log\frac{1-w}{w}\dff \eta.
\end{eqnarray}
Based on the single-photon approximation, $X = 1$ is detected if and only if the following is satisfied,
\begin{equation}\label{eq.LLR_M}
LLR=\sum_{m=1}^{M}\log \frac{\left(1-\gamma_t\right)G_0(z_m)+\gamma_tG_1(z_m)}{\left(1-\gamma_b\right)G_0(z_m)+\gamma_bG_1(z_m)}>\eta.
\end{equation}

The log-likelihood ratio in (\ref{eq.LLR_M}) can be simplified as follows. Note that we have \begin{equation}
\frac{G_1(z)}{G_0(z)}=e^{a_2z^2+a_1z+a_0},
\end{equation}
where parameters $a_0$, $a_1$, and $a_2$ are given by
\begin{eqnarray}
a_2&=&\frac{1}{2\sigma_0^2}-\frac{1}{2\left(\sigma_0^2+\sigma^2\right)}, \nonumber \\
a_1&=&\frac{Ae}{\sigma_0^2+\sigma^2}, \nonumber \\
a_0&=&\frac{1}{2}\log{\frac{\sigma_0^2}{\sigma_0^2+\sigma^2}}-\frac{A^2e^2}{2\left(\sigma_0^2+\sigma^2\right)}.
\end{eqnarray}
Letting $x_m = a_2z^2 + a_1z + a_0$ for $1 \leq m \leq M$, we have the following on the LLR,
\begin{eqnarray}\label{eq.LLR_F}
LLR=\sum_{m=1}^{M}\log\frac{(1-\gamma_t)+\gamma_te^{x_m}}{(1-\gamma_b)+\gamma_be^{x_m}}\dff \sum^M_{m=1}F(x_m).
\end{eqnarray}

Note that in the above summation for each $F(x_m)$, the numerator and the denominator contain the exponential term $e^{x_m}$. Such detector may suffer a high computational complexity since both the exponential and the logarithm operations are involved. To reduce such complexity, we resort to a piecewise approximation on $F(x)$ in terms of $x$, which is provided in the following subsection.
\subsection{The Piecewise Approximation for the MAP Detection}
We provide piecewise approximations on the function $F(x)$ to reduce the computational complexity. We first provide a piecewise linear approximation, which shows virtually the same detection error probability as that of the accurate LLR in the low thermal noise variance regime. To reduce the detection error probability in the high thermal noise variance regime, we further propose a piecewise polynomial approximation up to the cubic term.

We provide the following properties on the function $F(x)$, which will be used in the piecewise approximation.

\begin{theorem}
We have the following limit on $F(x)$ as $x$ approaches infinity,
\begin{eqnarray}
\lim_{x\rightarrow +\infty}F(x)&=&\log \frac{\gamma_t}{\gamma_b}, \label{equ.LimitX1}\nonumber \\
\lim_{x\rightarrow -\infty}F(x)&=&\log \frac{1-\gamma_t}{1-\gamma_b}. \label{equ.LimitX2}
\end{eqnarray}
Moreover, we have that $F(x)$ is central symmetric on the point $\left(x_0,F\left(x_0\right)\right)$, where $x_0$ is given by
\begin{eqnarray}\label{eq.x_0}
x_0&=&\frac{1}{2}\log \frac{(1-\gamma_t)(1-\gamma_b)}{\gamma_t\gamma_b}.
\end{eqnarray}
In other words, we have that $F(x_1) + F(x_2) = 2F(x_0)$ for $x_1 + x_2 = 2 x_0$.

\begin{proof}
Equations (\ref{equ.LimitX2}) is straightforward.
In the following we focus on the proof of (\ref{eq.x_0}).
To achieve this, we need to prove that $F(x_0)-F(x_0-x)=F(x_0+x)-F(x_0)$. Note that since
\begin{eqnarray}
F(x_1)-F(x_2)=\int_{x_2}^{x_1}F'(t)dt,
\end{eqnarray}
we need to prove that the following is satisfied for all $x$,
\begin{equation}\label{eq.condition2}
F'\left(x_0- x\right)=F'\left(x_0+x\right).
\end{equation}
Equation (\ref{eq.condition2}) can be proved via substituting $x_0 + x$ and $x_0 - x$ into the following,
\begin{equation}\label{eq.derivative_F}
F'(x)=\frac{\beta-\alpha}{\alpha\beta e^{x}+(\alpha+\beta-2\alpha\beta)+(1-\alpha)(1-\beta)e^x}.
\end{equation}
\end{proof}
\end{theorem}

Based on the above properties, we propose the piecewise linear approximation and piecewise cubic polynomial approximation on $F(x)$.
\subsubsection{Piecewise Linear Approximation}
The piecewise linear approximation is based on three lines, where two lines are the limits of $F(x)$ as $x \rightarrow +\infty$ and $x \rightarrow -\infty$, and one line is the tangent line of $F(x)$ at the point $x = x_0$. More specifically, the three-line piecewise function is given as follows,
\begin{equation}
F(x)\approx g(x)=\left\{\begin{array}{ll}
\log \frac{1-\gamma_t}{1-\gamma_b}, & x<x_1; \\
k\left(x-x_0\right)+F\left(x_0\right), & x_1\le x < x_2; \\
\log \frac{\gamma_t}{\gamma_b}, & x\ge x_2;
\end{array} \right.
\end{equation}
where slope $k = \frac{\partial F(x)}{\partial x}\Big|_{x = x_0}$, point $x_1=\frac{\log\frac{1-\gamma_t}{1-\gamma_b}-F(x_0)}{k}+x_0$, and point $x_2=\frac{\log \frac{\gamma_t}{\gamma_b}-F(x_0)}{k}+x_0$. As $k\rightarrow+\infty$, the three-line piecewise approximation degrades to a two-value step function.
\subsubsection{Piecewise Cubic Polynomial Approximation}
The entire range of variable $x$ is divided into five parts. The function $F(x)$ follows approximations using the limits as $x$ approaches positive infinity and negative infinity in the leftmost and the rightmost parts, respectively, and using cubic polynomial functions in the three middle parts. The five parts in the entire range of $x$ are given by $(-\infty, -x_0]$, $(-x_0, 0]$, $(0, x_0]$, $(x_0, 2x_0]$ and $(2x_0, +\infty)$. The piecewise approximation of $F(x)$ is given as follows,
\begin{eqnarray}\label{wpiececubic}
\begin{aligned}
F(x)\approx g(x)=\left\{\begin{array}{ll}
\log \frac{1-\gamma_t}{1-\gamma_b}, & x<-x_0; \\
k_1(4)x^3+k_1(3)x^2+k_1(2)x+k_1(1),& -x_0\le x < 0; \\
k_2(4)x^3+k_2(3)x^2+k_2(2)x+k_2(1), & 0 \le x<x_0; \\
k_3(4)x^3+k_3(3)x^2+k_3(2)x+k_3(1), & x_0 \le x<2x_0; \\
\log \frac{\gamma_t}{\gamma_b}, & x\ge 2x_0.
\end{array} \right.
\end{aligned}
\end{eqnarray}

Note that intuitively the divided parts for the piecewise approximation need to be symmetric in terms of the point $x_0$. However, according to the distribution of $z_m$, the probability of $x_m$ belonging to $(-\infty, x_0]$ is significantly larger than that of $x_m$ belonging to $[x_0, \infty)$. The approximation accuracy in the region $(-\infty, x_0]$ needs to be higher than that in the region $[x_0, \infty)$. Thus, we divide region $(-\infty, x_0]$ into more parts.

Let $\bk_i \dff [k_i(0)\quad k_i(1)\quad k_i(2)\quad k_i(3)]^T$ denote the coefficient vector of the cubic polynomial for $1 \leq i \leq 3$. Assume that in each part $i$, $J$ equally spaced samples $\{v_{ij}\}^J_{j=1}$ are employed for the approximation. Let $\bv_{ij} \dff [1\quad v_{ij}\quad v^2_{ij}\quad v^3_{ij}]^T$, the polynomial approximation is based on the following least squares criterion,
\begin{equation}
\min_\mathbf{\bk_i} \sum_{j=1}^{J}f(v_{ij})\left( \bv^T_{ij}\bk_i-F(v_{ij})\right)^2,
\end{equation}
where $f(v_{ij})$ denotes the probability density of $x$ at the point $x = v_{ij}$.
Here we adopt the probability density $f(v_{ij})$ as the weight for the approximation. The PDF of $x$ is given by,
\begin{eqnarray} \label{equ.PDFx}
f(x) \frac{dx}{dz} = p(z),
\end{eqnarray}
where $\frac{dx}{dz}$ can be derived according to the relationship $x = a_2z^2 + a_1z + z_0$.

Note that the polynomial approximation can be expressed by the following quadratic form,
\begin{eqnarray}
\min_{\bk_i}\bk^T_i\Big(\sum^J_{j=1}f(v_{ij})\bv_{ij}\bv^T_{ij}\Big)\bk_i - 2\Big(\sum^J_{j=1}f(v_{ij})F(v_{ij})\bv_{ij}\Big)\bk_{i} + \sum^J_{j=1}f(v_{ij})F^2(v_{ij}),
\end{eqnarray}
where the optimal solution is given by
\begin{equation}\label{eq.wcubic_k}
\bk_i=\Big(\sum_{j=1}^Jf(v_{ij})\bv_{ij}\bv^T_{ij}\Big)^{-1}\sum^J_{j=1}f(v_{ij})F(v_{ij})\bv_{ij}.
\end{equation}

Numerical results show that such a piecewise cubic polynomial approximation can significantly reduce the detection error probability compared with the piecewise linear approximation, especially for a large thermal noise variance.

\section{Non-ideal Photon-counting Receiver}
The ideal photo-counting receiver that counts the number of detected photoelectrons is difficult to implement in practice. Instead, a non-ideal the photo-counting receiver is implemented based on a WP receiver. A hard decision is employed for the PMT output signal within each interval, where a photoelectron is detected if the PMT output is larger than a certain threshold. A crucial question involved is the selection of the optimal threshold, to minimize the detection error probability. To solve this, we propose the threshold selection based on the exact detection error probability and the KL distance.
\subsection{The Signal Model for a Non-ideal Photon-counting Receiver}
Recall that the PMT output signals in the $M$ intervals are given by $\bz = [z_1, z_2, ..., z_M]^T$. Let $\bb = [b_1, b_2, ..., b_M]$ denote indicators on whether a photoelectron is detected in each interval. A threshold $z_{th}$ is employed for the detection, where each indicator $b_m$ is given by
\begin{eqnarray}
b_m=\left\{\begin{array}{ll}
0, & z_m<z_{th}; \\
1, & z_m\ge z_{th}.
\end{array} \right.
\end{eqnarray}

Note that for OOK symbols $X = 0$ and $X = 1$, each component of the corresponding indicator vector $\bb$ satisfies the corresponding binary distributions. The optimal detection between the two vectors is based on the summation of each component of $\bb$.
For symbols $X=1$ and $X=0$, the probabilities of each component of $\bb$ being one, denoted as $p_1$ and $p_0$, respectively, are given by
\begin{eqnarray}
p_1&=&\left(1-\gamma_t\right)Q\left(\frac{z_{th}}{\sigma_0}\right)+\gamma_tQ\left(\frac{z_{th}-Ae}{\sqrt{\sigma_0^2+\sigma^2}}\right), \nonumber \\
p_0&=&\left(1-\gamma_b\right)Q\left(\frac{z_{th}}{\sigma_0}\right)+\gamma_bQ\left(\frac{z_{th}-Ae}{\sqrt{\sigma_0^2+\sigma^2}}\right).
\end{eqnarray}
Let $B_M = \sum^M_{m=1}b_m$ be the summation of the $M$ indicators. We have that for $X = i$, $i = 0,1$, the summation $B_M$ satisfies the following binomial distributions
\begin{eqnarray} \label{equ.Binomial}
\mathbb{P}(B_M = n) = {M \choose n}p^n_i(1-p_i)^{M-i}.
\end{eqnarray}

\subsection{The Threshold Selection Rules}
The optimal detection for such two different binomial distributions are based on the threshold detection on the summation $B_M$. Note that $p_1$ and $p_0$ are strictly decreasing with respect to $z_{th}$. A natural question is to select the optimal threshold $z_{th}$ that aims to minimize the detection error probability. This can be solved based on two criteria, to minimize the exact detection error probability and to maximize the KL distance between two Bernoulli distributions. As the exact error probability-based criterion might be intractable due to the high computational complexity, the KL distance-based criterion can serve as an low computational complexity solution.

For the exact detection error probability-based criterion, symbol $X = 1$ is detected if  and only if
\be
\log\frac{\mathbb{P}(B_M = n|X=1)}{\mathbb{P}(B_M = n|X=0)}>\eta = \log\frac{1-w}{w}.
\ee
We can obtain a detection threshold $B_{th}$ such that $X=1$ is detected if $B_M>B_{th}$ and $X = 0$ is detected otherwise, where the threshold is given by
\begin{equation}\label{sub_th}
B_{th}=\left\lfloor  \frac{\eta-M\log\frac{1-p_1}{1-p_0}}{\log\frac{p_1}{p_0}-\log\frac{1-p_1}{1-p_0}}\right\rfloor.
\end{equation}
The probabilities of symbol $0$ being detected for $X = 1$ and symbol $1$ being detected for $X = 0$,  denoted as $p^e_{01}$ and $p^e_{10}$, respectively, are given by
\begin{eqnarray}\label{p_error}
p^e_{01}&=&\sum_{k=0}^{k\leq B_{th}}\binom{M}{k}p_1^k(1-p_1)^{M-k}, \\
p^e_{10}&=&\sum_{k>B_{th}}^{\infty}\binom{M}{k}p_0^k(1-p_0)^{M-k}.
\end{eqnarray}
The optimal threshold, denoted as $\hat z_{th}$, can be selected such that the detection error probability $(1-w)p^e_{01} + wp^e_{10}$ is minimized. More specifically, the optimal threshold $\hat z_{th}$ is given by
\begin{eqnarray}
\hat z_{th} = \arg\min_{z_{th}}(1-w)p^e_{01} + wp^e_{10}.
\end{eqnarray}

The KL distance-based criterion is based on the method of types, where the two binomial distributions correspond to two types according to the Chernoff-Stein Lemma \cite{ElementsInfoTheory}. As the number of intervals $M$ approaches infinity, we have the following on the exponents of $p^e_{01}$ and $p^e_{10}$,
\begin{eqnarray}\label{error_type}
-\frac{1}{M}\log p^e_{01}&\rightarrow& D\left(p_0||p_1\right)\dff p_0\log\frac{p_0}{p_1}+(1-p_0)\log\frac{1-p_0}{1-p_1}, \\
-\frac{1}{M}\log p^e_{10}&\rightarrow& D\left(p_1||p_0\right)\dff p_1\log\frac{p_1}{p_0}+(1-p_1)\log\frac{1-p_1}{1-p_0}.
\end{eqnarray}
The optimal threshold $z_{th}$ is selected to maximize the minimum KL distance. More specifically, the optimal threshold $\hat z_{th}$ is given by
\begin{eqnarray}
\hat z_{th} = \arg\max_{z_{th}}\min\Big\{D\left(p_0||p_1\right), D\left(p_1||p_0\right)\Big\}.
\end{eqnarray}

\subsection{Sensitivity Analysis for the Detection Threshold}
We focus on the optimal detection threshold that maximizes the minimum KL distance. We provide the upper and lower bounds on the optimal detection threshold, and further prove that for sufficiently small $\sigma^2$ and $\sigma^2_0$, a slight deviation from the optimal detection threshold does not significantly decrease the minimum KL distance. This result shows that the minimum KL distance is not sensitive to the detection threshold selection in a practical system.

To prove this, we first obtain the following upper and lower bounds on the optimal detection threshold $\hat z_{th}$.

\begin{theorem}
We have the upper bound $Z(C_u)$ and lower bound $Z(C_l)$ on the optimal threshold, where parameters $C_l$ and $C_u$ are independent of $\sigma_0$ and $\sigma$, and
\be
Z(C)=\frac{-\frac{Ae}{\sigma^2+\sigma_0^2}+\sqrt{\frac{A^2e^2}{(\sigma^2+\sigma_0^2)^2}-\left(\frac{1}{\sigma^2+\sigma_0^2}-\frac{1}{\sigma_0^2}\right)\left(\frac{A^2e^2}{\sigma^2+\sigma_0^2}-2\log\frac{C\sigma_0}{\sqrt{\sigma^2+\sigma_0^2}}\right)}}{\frac{1}{\sigma_0^2}-\frac{1}{\sigma^2+\sigma_0^2}};
\ee

\begin{proof}
Please refer to Appendix A.
\end{proof}
\end{theorem}

Based on Theorem~$6$, we have the following bounds on the optimal detection threshold $\hat z_{th}$.

\begin{theorem}
For any $\epsilon > 0$, we have that for sufficiently small  $\sigma_0$, the optimal detection threshold $\hat{z}_{th}$ satisfies
\be
\sigma_0\left(\sqrt{-2(1-\epsilon\sigma_0)\log\sigma_0}-\epsilon Ae\right)< \hat{z}_{th}< \frac{Ae}{2}+\epsilon.
\ee
\begin{proof}
First we prove the upper bound. Note that
\be
\hat{z}_{th}&<&\frac{-\frac{Ae}{\sigma^2+\sigma_0^2}+\sqrt{\frac{A^2e^2}{(\sigma^2+\sigma_0^2)^2}-\left(\frac{1}{\sigma^2+\sigma_0^2}-\frac{1}{\sigma_0^2}\right)\left(\frac{A^2e^2}{\sigma^2+\sigma_0^2}-2\log\frac{C_u\sigma_0}{\sqrt{\sigma^2+\sigma_0^2}}\right)}}{\frac{1}{\sigma_0^2}-\frac{1}{\sigma^2+\sigma_0^2}} \nonumber \\
&=&\frac{-\frac{Ae}{\sigma^2+\sigma_0^2}+\sqrt{\frac{A^2e^2}{(\sigma^2+\sigma_0^2)\sigma^2_0}+\left(\frac{1}{\sigma^2+\sigma_0^2}-\frac{1}{\sigma_0^2}\right)\left(2\log \frac{C_u\sigma_0}{\sqrt{\sigma^2 + \sigma^2_0}}\right)}}{\frac{1}{\sigma_0^2}-\frac{1}{\sigma^2+\sigma_0^2}} \label{theorem7.right1}\\
&<&\frac{-\frac{Ae}{\sigma^2+\sigma_0^2}+\frac{Ae}{\sqrt{\left(\sigma_0^2+\sigma^2\right)\sigma_0^2}}+\frac{\sqrt{\left(\sigma_0^2+\sigma^2\right)\sigma_0^2}}{Ae}\left(\frac{1}{\sigma_0^2}-\frac{1}{\sigma^2+\sigma_0^2}\right)\left|\log\frac{C_u\sigma_0}{\sqrt{\sigma^2 + \sigma^2_0}}\right|}{\frac{1}{\sigma_0^2}-\frac{1}{\sigma^2+\sigma_0^2}} \\
&=&Ae\frac{\frac{1}{\sqrt{\sigma^2+\sigma_0^2}}}{\frac{1}{\sigma_0}+\frac{1}{\sqrt{\sigma^2+\sigma_0^2}}}+\frac{\sqrt{\left(\sigma_0^2+\sigma^2\right)\sigma_0^2}}{Ae}\left|\log\frac{C_u\sigma_0}{\sqrt{\sigma^2 + \sigma^2_0}}\right|
 \\
&\le&\frac{Ae}{2}+\frac{\sqrt{\left(\sigma_0^2+\sigma^2\right)\sigma_0^2}}{Ae}\left|\log\frac{C_u\sigma_0}{\sqrt{\sigma^2 + \sigma^2_0}}\right|.\label{theorem7.right2}
\ee
Note that for the upper bound given in (\ref{theorem7.right2}), we have that
\be
\lim_{\sigma_0\rightarrow0}\frac{Ae}{2}+\frac{\sqrt{\left(\sigma_0^2+\sigma^2\right)\sigma_0^2}}{Ae}\left|\log\frac{C_u\sigma_0}{\sqrt{\sigma^2 + \sigma^2_0}}\right|=\frac{Ae}{2}
\ee
Then for any $\epsilon > 0$, we have that for sufficiently small $\sigma_0$, $\hat z_{th} < \frac{Ae}{2} + \epsilon$.

Next we prove the lower bound. We have that
\be
\hat{z}_{th}&>&\frac{-\frac{Ae}{\sigma^2+\sigma_0^2}+\sqrt{\frac{A^2e^2}{(\sigma^2+\sigma_0^2)\sigma^2_0}+\left(\frac{1}{\sigma^2+\sigma_0^2}-\frac{1}{\sigma_0^2}\right)\left(2\log \frac{C_l\sigma_0}{\sqrt{\sigma^2 + \sigma^2_0}}\right)}}{\frac{1}{\sigma_0^2}-\frac{1}{\sigma^2+\sigma_0^2}}
\nonumber \\
&>&\frac{-\frac{Ae}{\sigma^2+\sigma_0^2}+\sqrt{\frac{A^2e^2}{(\sigma^2 + \sigma^2_0)\sigma^2_0} + \left(\frac{1}{\sigma_0^2}-\frac{1}{\sigma^2+\sigma_0^2}\right)\left(-2\log\frac{C_{l}\sigma_0}{\sqrt{\sigma^2+\sigma_0^2}}\right)}}{\frac{1}{\sigma_0^2}}.
 \label{theorem7.left1}\\
&=&-\frac{Ae\sigma_0^2}{\sigma^2+\sigma_0^2}+\sigma_0\sqrt{\frac{A^2e^2}{(\sigma^2 + \sigma^2_0)} + \left(1-\frac{\sigma_0^2}{\sigma^2+\sigma_0^2}\right)\left(-2\log\frac{C_{l}\sigma_0}{\sqrt{\sigma^2+\sigma_0^2}}\right)}. \label{theorem7.left2}
\ee
Given any $\sigma > 0$, we consider sufficient small $\sigma_0$, such that the following inequality is satisfied based on (\ref{theorem7.left2}),
\be
\hat{z}_{th}&>&-\frac{Ae\sigma_0^2}{\sigma^2+\sigma_0^2}+\sigma_0\sqrt{\left(1-\frac{\sigma_0^2}{\sigma^2+\sigma_0^2}\right)\left(-2\log\frac{C_{l}\sigma_0}{\sqrt{\sigma^2+\sigma_0^2}}\right)} \nonumber \\
&\ge& \sigma_0\left(\sqrt{-2(1-\epsilon\sigma_0)\log\sigma_0}-\epsilon Ae\right).
\ee
\end{proof}
\end{theorem}

Based on Theorem~$7$, the following results shows that the crossover probabilities
$p_0$ and $p_1$ are not sensitive to the detection threshold $z_{th}$.

\begin{theorem}
For any $\delta_0 > 0$ and $\epsilon > 0$, there exists a sufficiently small $\sigma$ and $\kappa>0$ such that for $\sigma_0 < \kappa$, we have that $|p_0-\gamma_b|<\epsilon$ and $|p_1-\gamma_t|<\epsilon$ for any detection threshold $z_{th} \in [\hat z_{th}, Ae - \delta_0)$.

\begin{proof}
For any $\delta_0 > 0$ and $\epsilon > 0$, we have the following for $z_{th} < Ae - \delta_0$,
\be
\frac{Ae-z_{th}}{\sqrt{\sigma^2+\sigma_0^2}}&\ge&\frac{\delta_0}{\sqrt{\sigma^2+\sigma_0^2}}.
\ee
Letting $\sigma = \frac{\delta_0}{\sqrt{2}Q^{-1}(\epsilon)}$, we have that for $\sigma_0 < \sigma$,
\be
Q\left(\frac{Ae - z_{th}}{\sqrt{\sigma^2 + \sigma^2_0}}\right) < \epsilon, \ \ \mbox{for} \ \ z_{th} < Ae - \delta \ \ \mbox{and} \ \ \sigma_0 < \sigma.
\ee
According to Theorem~$7$, we have that there exists $\kappa_0 > 0$ such that for any $\sigma_0 < \kappa_0$,
\be
\hat z_{th} \geq \sigma_0\left(\sqrt{-2(1-\epsilon\sigma_0)\log\sigma_0}-\epsilon Ae\right) > \sigma_0 Q^{-1}(\epsilon).
\ee
Consider $\sigma_0 < \kappa \dff \min\{\xi_0, \sigma\}$. Then for $\hat z_{th} < z_{th} < Ae - \delta_0$, we have
\be
|p_0-\gamma_b|=\left|(1-\gamma_b)Q\left(\frac{z_{th}}{\sigma_0}\right)+\gamma_bQ\left(\frac{Ae-z_{th}}{\sqrt{\sigma^2+\sigma_0^2}}\right)\right|<(1-\gamma_b)Q\left(Q^{-1}(\epsilon)\right)+\gamma_bQ\left(Q^{-1}(\epsilon)\right)=\epsilon.
\ee
The same procedure can be performed to prove that $|p_1 - \gamma_t| < \epsilon$.
\end{proof}
\end{theorem}

Based on the above result, since $p_0$ and $p_1$ are not sensitive to the detection threshold $z_{th}$, the KL distance is also not sensitive. It implies that for sufficient small thermal noise variance $\sigma_0$ and shot noise variance $\sigma$, the detection threshold $z_{th}$ can be selected between a value larger than $\frac{Ae}{2}$ and a value smaller than $Ae$, without substantial performance loss in terms of the KL distances. Such insights show significant values for the detection threshold selection in a practical scattering communication system.
\section{Numerical Results}
Consider a WP receiver with additive shot noise and thermal noise. Assume that for the shot noise variance $\sigma^2=\xi^2A^2e^2$, the PMT spreading factor $\xi=0.1$. Define $SNR=\log_{10} \frac{\gamma_s}{\gamma_b}$ as the ratio between the signal intensity and the background radiation intensity.

Figure~\ref{fig.capacity_single_approx} compares the transmission rate from the single-photon approximation and the true rate from the Poisson distribution for different thermal noise variance, for the $\gamma_t = 0.01$ and $0.05$.
It is seen that the rate from the signal-photon approximation can well approximate the true rate for $\gamma_t = 0.01$.
The same approximation performance of the single-photon approximation can be observed for $\gamma_t = 0.02$.
However, the gap between the true rate and the approximation becomes non-negligible for $\gamma_t = 0.05$ and $\sigma_0 = 0.2Ae$.
It is seen that the single-photon approximation works well for the small photon probability up to $\gamma_t = 0.02$.

Figure~\ref{scalar_bounds} shows the upper and lower bounds on the rate against the SNR for the thermal noise variance $\sigma_0^2 = 0.1^2A^2e^2$, $0.15A^2e^2$, and $0.2^2A^2e^2$. It is seen that the gap between the two bounds becomes negligible for the thermal noise variance $\sigma_0\le0.1^2A^2e^2$, i.e., the PMT amplification factor $A\ge10\frac{\sigma_0}{e}$. Moreover, Figure~\ref{vector_bounds} shows the upper and lower bounds on the rate against the number of intervals $M$ for the $SNR=20dB$ and the thermal noise variance $\sigma_0^2 = 0.1^2A^2e^2$, $0.15A^2e^2$, and $0.2^2A^2e^2$. Again, it is seen that gap between the two bounds becomes negligible for the thermal noise variance $\sigma_0^2\le0.1^2A^2e^2$.

Assume that the number of intervals $M = 1000$. We compare the piecewise linear approximation and cubic approximation result in Figure~\ref{fig.fitting} for the $SNR = 20dB$. It is seen that the gap between true function and cubic approximation function is negligible, while the gap between true function and linear approximation function is significant. Figure~\ref{fig.approx_error} shows the error performance against thermal noise variance $\sigma_0$. It is seen that the detection error for proposed two approximations becomes negligible for the thermal noise variance $\sigma_0^2\le0.1^2A^2e^2$. However, the performance loss for the linear approximation becomes non-negligible for the thermal variance $\sigma_0^2\ge0.15^2A^2e^2$, while the performance loss for the piecewise cubic  polynomial approximation is still negligible for the thermal variance $\sigma_0^2\ge0.2^2A^2e^2$.
This can be justified by the fact that for larger thermal noise variance, there is a larger probability of $x = a_2z^2 + a_1z + a_0$
falling into the region with non-negligible gap for the piecewise linear approximation.
\begin{figure}[htb]
\centering
\includegraphics[width = 0.8\columnwidth]{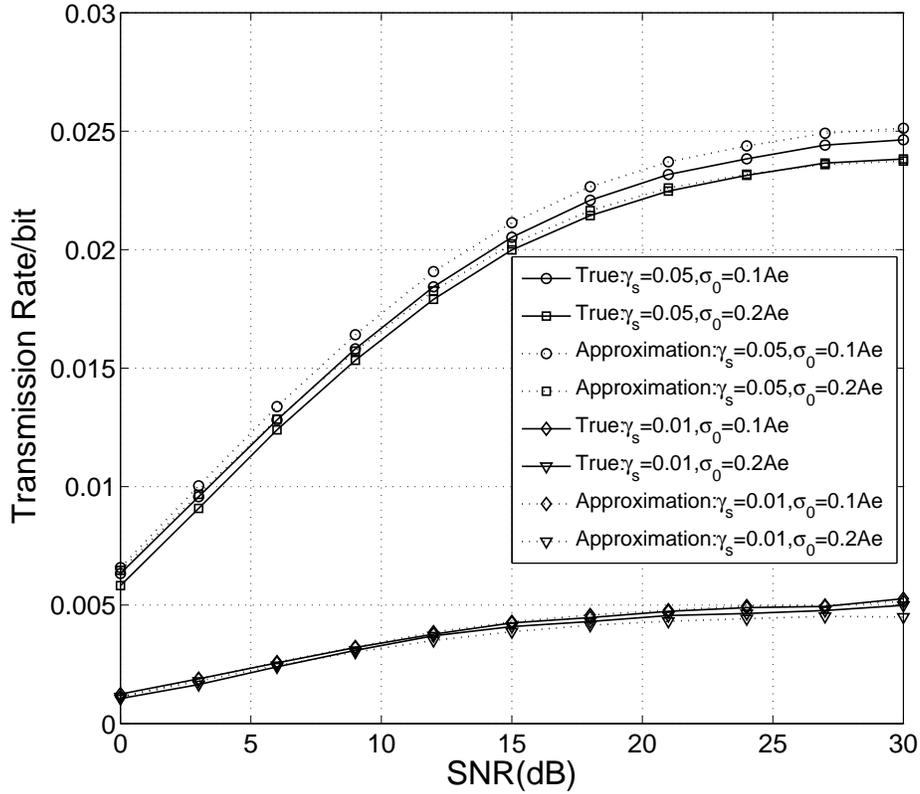}
\caption{The true rate and the rate from the single-photon approximation. }
\label{fig.capacity_single_approx}
\end{figure}
\begin{figure}[htb]
\centering
\includegraphics[width = 0.8\columnwidth]{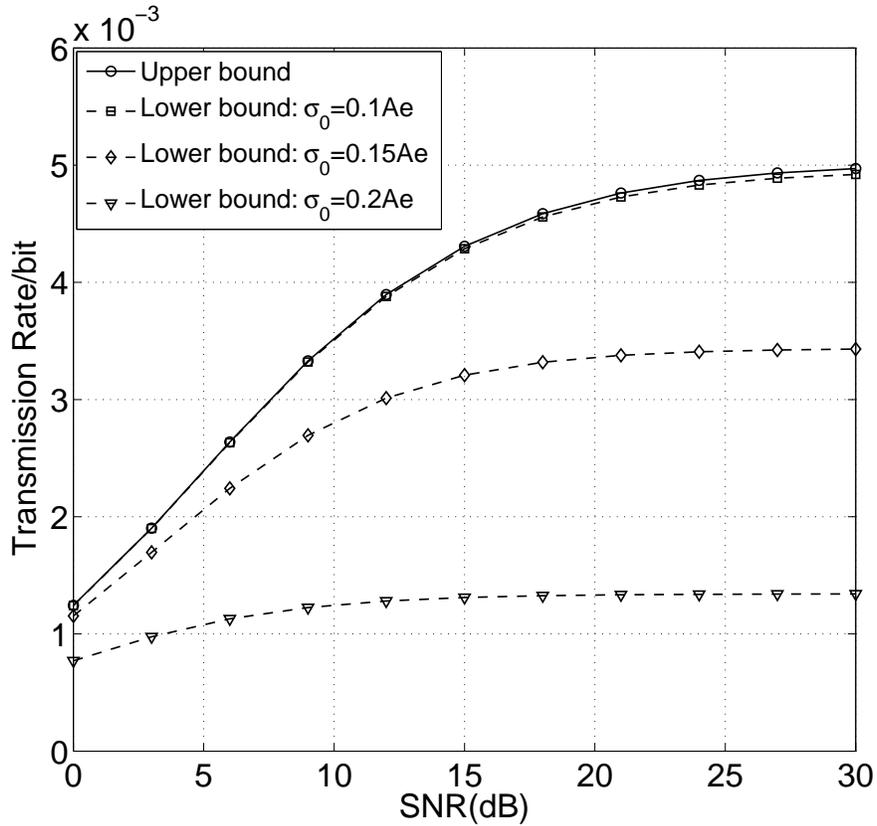}
\caption{The upper and lower bounds on the rate against SNR for the single PMT output interval.}
\label{scalar_bounds}
\end{figure}
\begin{figure}[htb]
\centering
\includegraphics[width = 0.8\columnwidth]{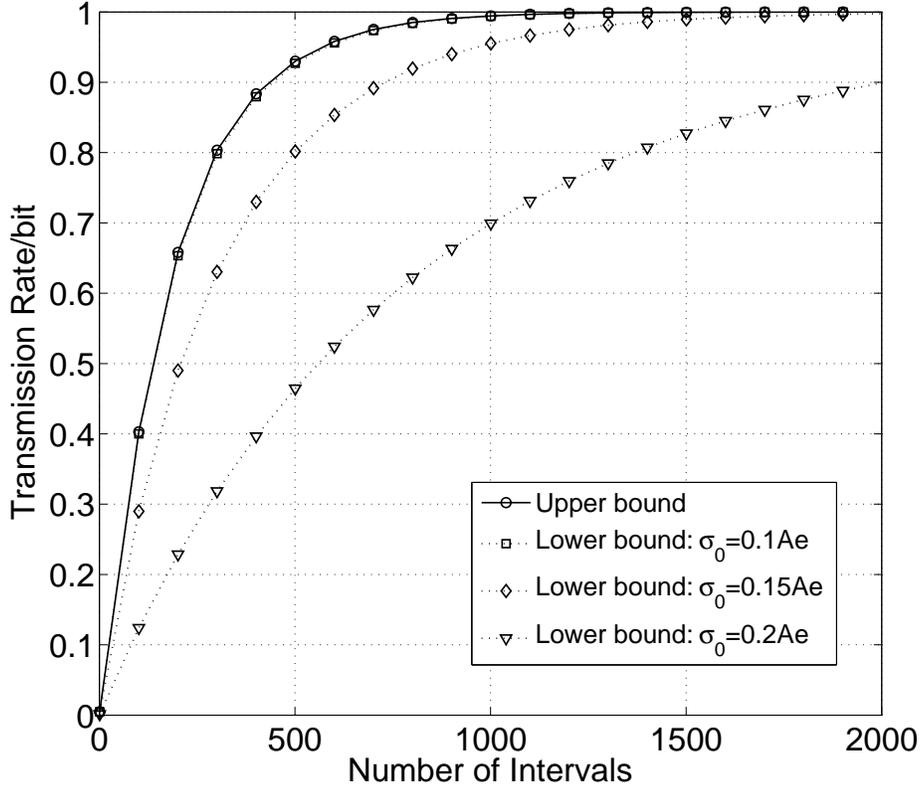}
\caption{The upper and lower bounds on the rate as a function of number of intervals for fixed $SNR=20dB$.}
\label{vector_bounds}
\end{figure}
\begin{figure}[htbp]
\centering
{\includegraphics[width = 0.8\columnwidth]{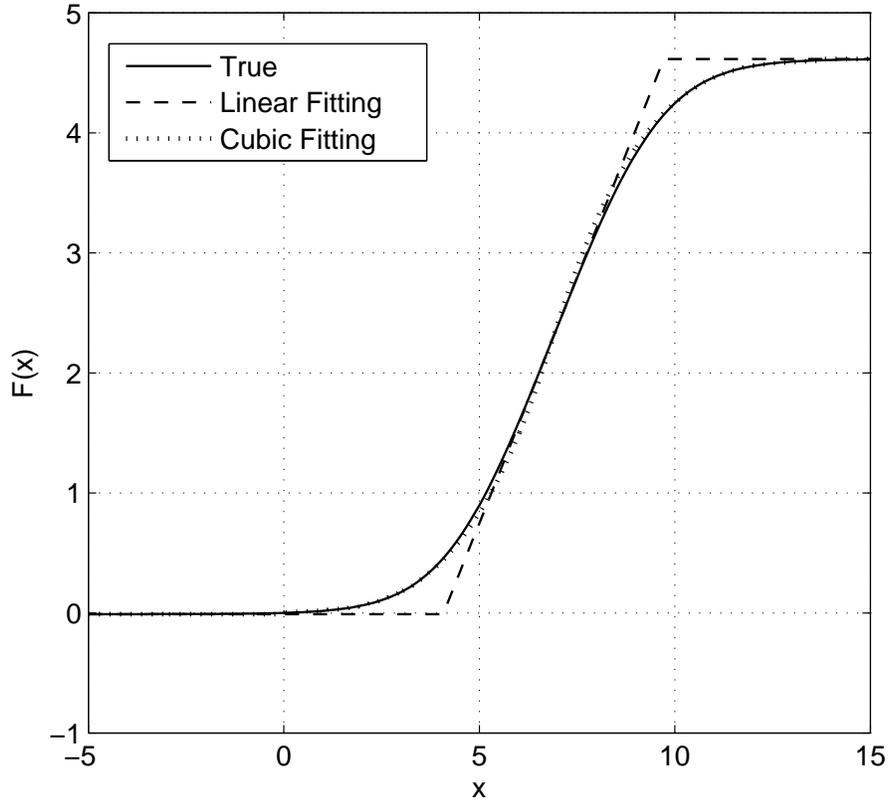}}
\caption{Piecewise linear and cubic polynomial approximations for function $F(x)$.}
\label{fig.fitting}
\end{figure}

\begin{figure}[htb]
\centering
{\includegraphics[width = 0.8\columnwidth]{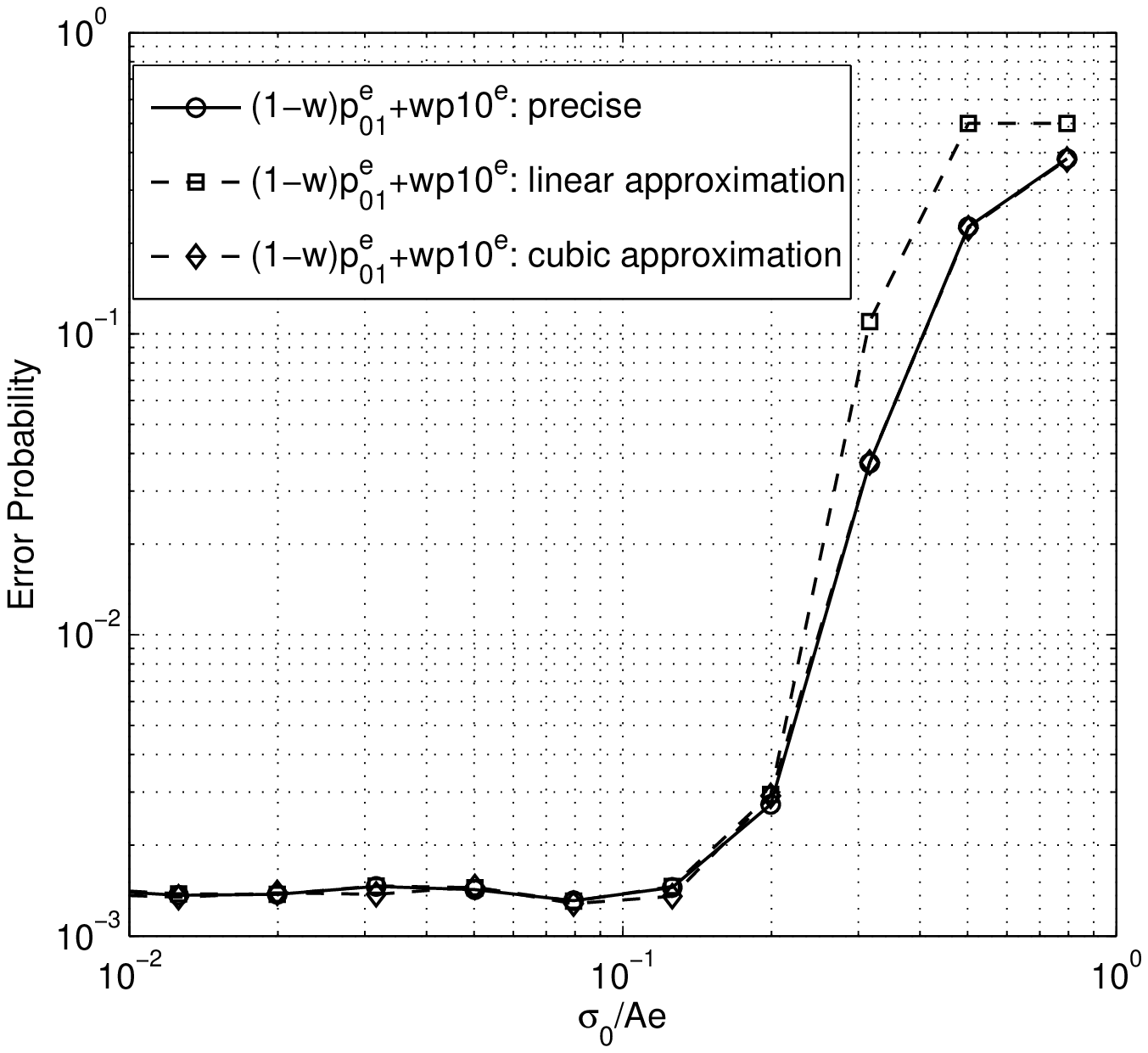}}
\caption{The detection error probability for the piecewise linear and cubic polynomial approximations.}
\label{fig.approx_error}
\end{figure}

For the non-ideal photon-counting receiver, assume that the symbol duration is divided into $M=1000$ slots, and that $SNR=20dB$. The optimal thresholds based on minimizing the detection error probability and maximizing the minimum KL distance are shown in Figure~\ref{threshold_KL_simulation}. It is seen that the optimal thresholds based on the true detection error probability and the corresponding KL distance are close to each other. Figure~\ref{errorcompare_binary} compares the detection error probability for the optimal threshold selection based on the exact detection error and the KL distance. It can be seen that threshold selection rule based on the KL distance shows negligible performance loss compared with that based on the exact detection error probability. Moreover, comparing Figure~$5$ and Figure~$7$, we have that the performance loss for the hard-decision of the non-ideal photon-counting receiver is negligible, compared with the optimal LLR based detection.
This shows that the non-ideal photon-counting receiver can serve as a good approximation to the optimal receiver,
with significantly reduced computational complexity.

Finally, we show the KL distance $\min\{D(p_0||p_1),D(p_1||p_0)\}$ with respect to the detection threshold $z_{th}$ in Figure~\ref{KLdistance_threshold} for the values of $(\sigma_0, \sigma) = (0.1Ae, 0.1Ae)$, $(0.05Ae,0.05Ae)$ and $(0.02Ae,0.02Ae)$. It is seen that for smaller noise variance, KL distance curve shows larger flat regime including the optimal detection threshold, where the KL distance is less sensitive to the detection threshold $z_{th}$. In other words, the small deviation of the detection threshold $z_{th}$ from the optimal one does not cause substantial performance loss in terms of the KL distance, which validates the results of Theorem~$8$.
\begin{figure}[htb]
\centering
\includegraphics[width = 0.8\columnwidth]{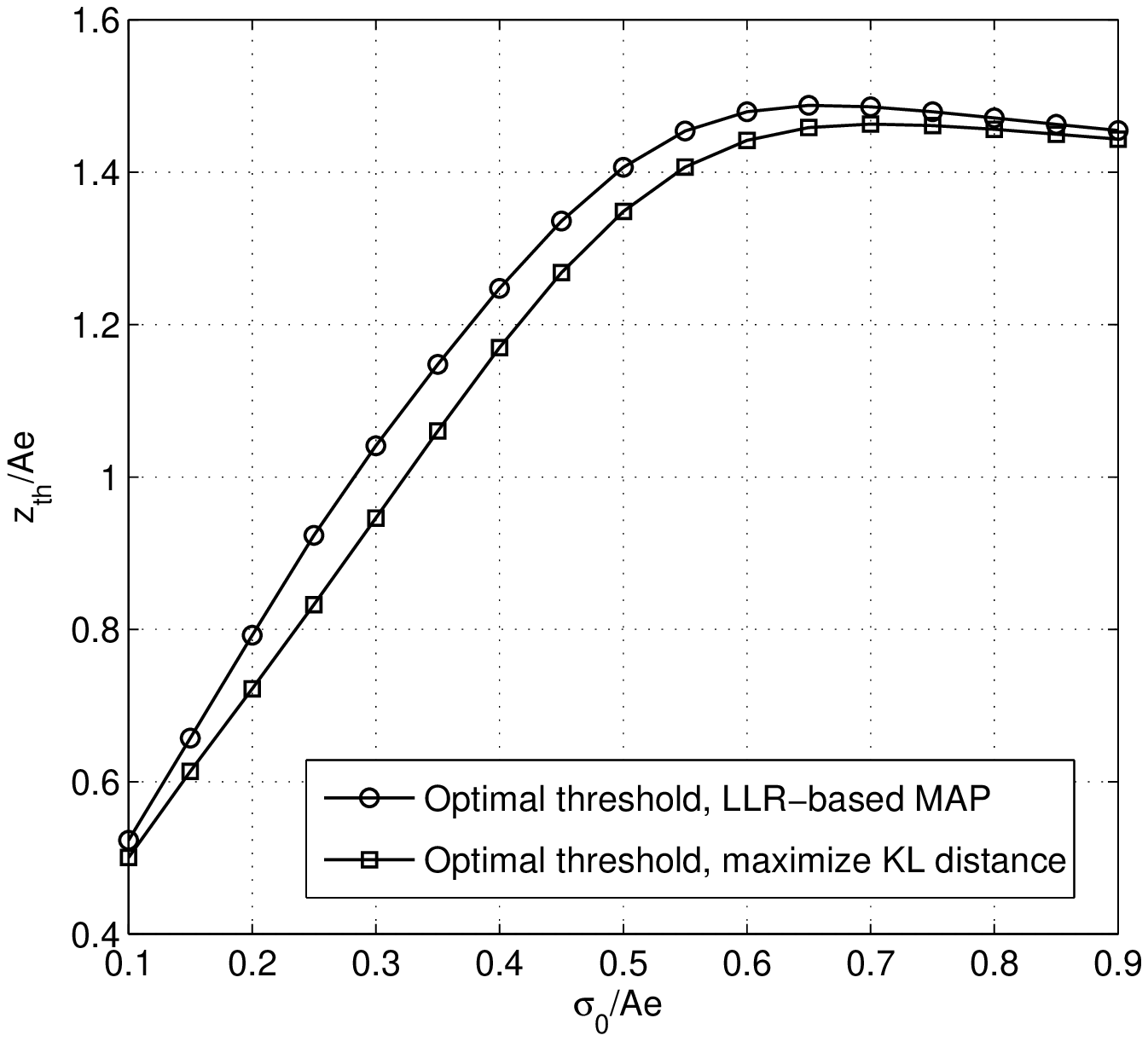}
\caption{The detection thresholds based on the true detection error probability and the KL distance, for the $SNR=20dB$ and number of intervals $M=1000$.}
\label{threshold_KL_simulation}
\end{figure}
\begin{figure}[htb]
\centering
\includegraphics[width = 0.8\columnwidth]{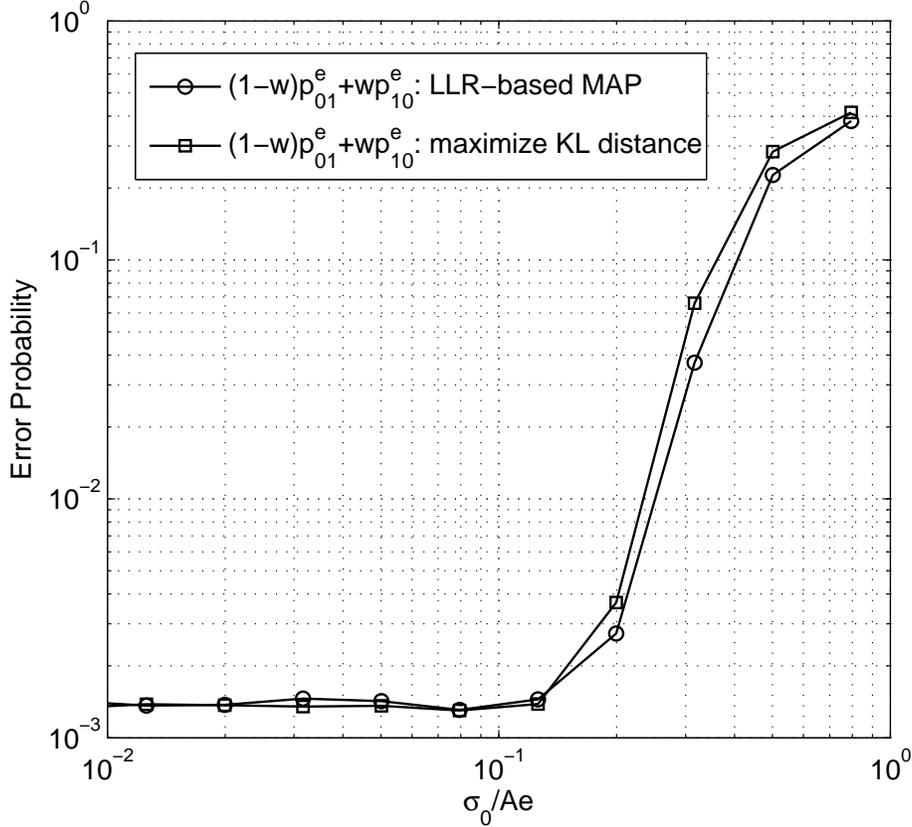}
\caption{The detection error probability for the threshold selection based on the exact detection error probability and the KL distance for the $SNR=20dB$ and number of intervals $M=1000$.}
\label{errorcompare_binary}
\end{figure}
\begin{figure}[htb]
\centering
\includegraphics[width = 0.8\columnwidth]{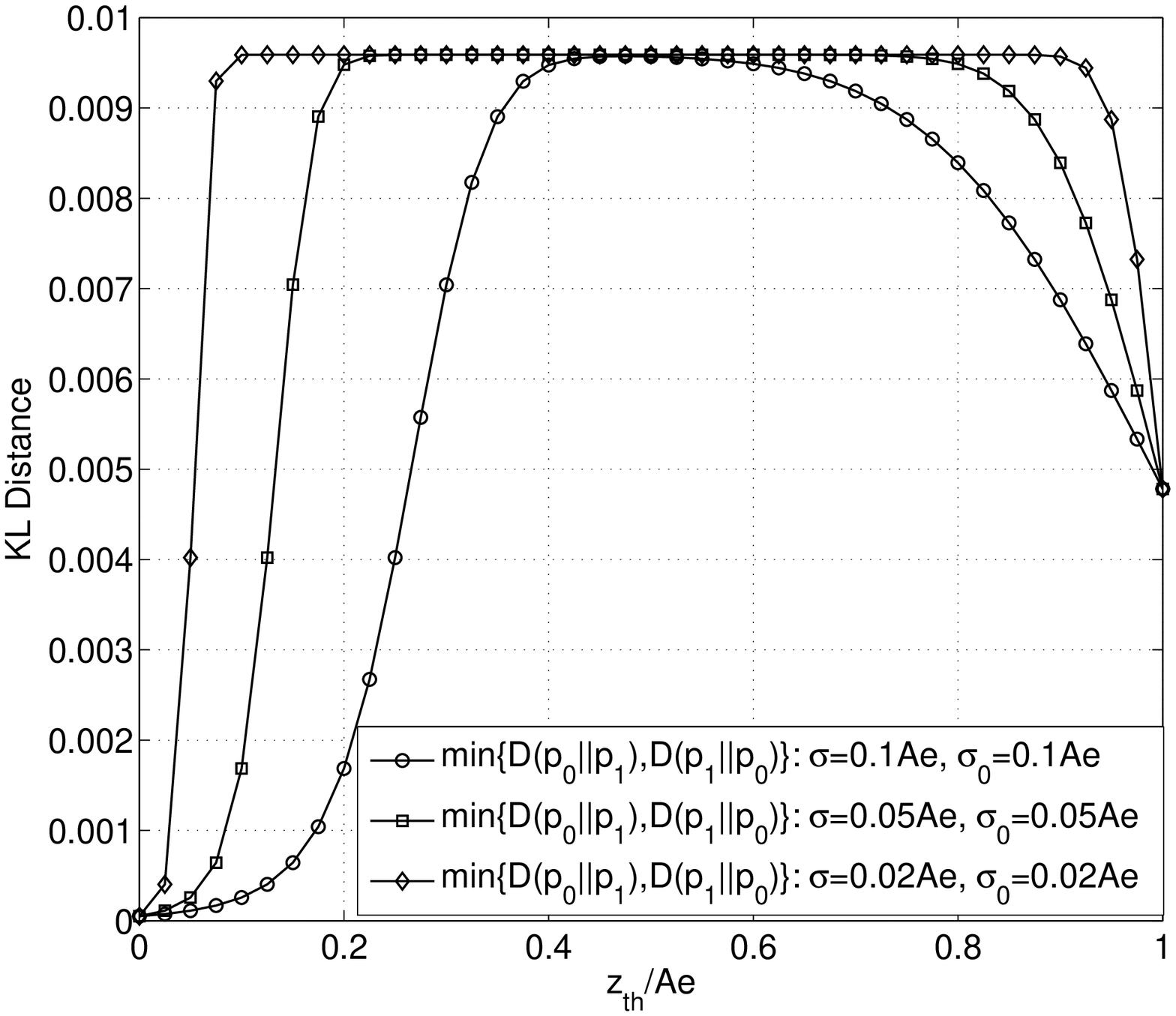}
\caption{The KL distance $D(p_0||p_1)$ and $D(p_1||p_0)$ versus threshold $z_{th}$ for different $\sigma$ and $\sigma_0$, for the $SNR=20dB$.}
\label{KLdistance_threshold}
\end{figure}
\section{Conclusions}
We have modeled the PMT output signal within each symbol duration as a vector of sampled analog signals within small intervals. Based on such a model, we have derived the upper and lower bounds on the transmission rate, which are proved to be tight for the PMT with a large amplification factor and small spreading factor.
We have investigated the MAP detector, and the reduced complexity receivers based on the piecewise linear and polynomial approximation. We have also presented a model for the non-ideal photon-counting receiver based on the hard-decision for the PMT output signals. The optimal threshold based on the KL distance is shown to be close to that based on the true detection error probability, with lower computational complexity.
Moreover, it is concluded that the non-ideal photon-counting receiver can serve as a good approximation to the optimal receiver,
with significantly reduced computational complexity, and the performance is not sensitive to the detection threshold selection for small thermal and shot noise variance.

\appendix
\subsection{Proof of Theorem 6}
We first investigate the detection threshold, denoted as $\hat z^{(1)}_{th}$, that maximizes the KL distance $D(p_0||p_1)$. Note that we have
\be\label{eq.opt1}
\frac{\partial D(p_0||p_1)}{\partial z_{th}}&=&\frac{\partial D(p_0||p_1)}{\partial p_0}\frac{\partial p_0}{\partial z_{th}}+\frac{\partial D(p_0||p_1)}{\partial p_1}\frac{\partial p_1}{\partial z_{th}} \nonumber \\
&=&\left(\log\frac{p_0}{p_1}-\log\frac{1-p_0}{1-p_1}\right)\frac{\partial p_0}{\partial z_{th}}+\left(\frac{1-p_0}{1-p_1}-\frac{p_0}{p_1}\right)\frac{\partial p_1}{\partial z_{th}}.
\ee
Then at the optimal threshold $\hat z^{(1)}_{th}$, we have the following,
\be\label{eq.opt2}
\frac{\log\frac{1-p_0}{1-p_1}-\log\frac{p_0}{p_1}}{\frac{1-p_0}{1-p_1}-\frac{p_0}{p_1}}=\frac{\frac{\partial p_1}{\partial z_{th}}}{\frac{\partial p_0}{\partial z_{th}}}.
\ee
where
\be\label{eq.opt3}
\frac{\partial p_0}{\partial z_{th}}&=&(1-\gamma_b)G(z_{th},0,\sigma_0^2)+\gamma_bG(z_{th},Ae,\sigma^2+\sigma_0^2) , \nonumber \\
\frac{\partial p_1}{\partial z_{th}}&=&(1-\gamma_t)G(z_{th},0,\sigma_0^2)+\gamma_tG(z_{th},Ae,\sigma^2+\sigma_0^2).
\ee
To simplify the following analysis, we define
$G_0\dff G(z_{th},0,\sigma_0^2)$,
$G_1\dff G(z_{th},Ae,\sigma^2+\sigma_0^2)$,
$Q_0\dff Q\left(\frac{z_{th}}{\sigma_0}\right)$, and
$Q_1\dff Q\left(\frac{z_{th}-Ae}{\sqrt{\sigma^2+\sigma_0^2}}\right)$.

Note that for $0 < z_{th} < Ae$, we have $Q_0 < \frac{1}{2} < Q_1$. Since $\gamma_b < \gamma_t$, we have the following
\be
p_0&=&(1-\gamma_b)Q_0+\gamma_bQ_1=Q_0+(Q_1-Q_0)\gamma_b\nonumber \\
&<&Q_0+(Q_1-Q_0)\gamma_t=p_1.
\ee

Based on the above equation and the definition on $p_0$ and $p_1$ [c.f. (61)], we have the following,
\be
\frac{1-p_0}{1-p_1}>1,\ \ \frac{p_0}{p_1}> \frac{\gamma_b}{\gamma_t} .
\ee

To bound the optimal threshold $\hat z^{(1)}_{th}$, we consider function $\frac{\log a-\log b}{a-b}$, which is shown to be decreasing with respect to $a$ and $b$ (the proof is given in Appendix.B). Then we have
\be\label{eq.threshold_upper}
\frac{\frac{\partial p_1}{\partial z_{th}}}{\frac{\partial p_0}{\partial z_{th}}}=\frac{\log\frac{1-p_0}{1-p_1}-\log\frac{p_0}{p_1}}{\frac{1-p_0}{1-p_1}-\frac{p_0}{p_1}}<\frac{1-\log\frac{\gamma_b}{\gamma_t}}{1-\frac{\gamma_b}{\gamma_t}}\dff C_1.
\ee
Substituting (\ref{eq.opt3}) into the above inequality, we have that for the $\hat z^{(1)}_{th}$ where $\frac{\partial D(p_0||p_1)}{\partial z_{th}} = 0$, the following is satisfied,
\be
\frac{G_0}{G_1}>\frac{\gamma_t-C_1\gamma_b}{C_1(1-\gamma_b)+\gamma_t-1}\dff C_{u0},\label{eq.threshold_upper2}
\ee
Considering the range $0 \leq z_{th} \leq Ae$ on the detection threshold, via directly solving (\ref{eq.threshold_upper2}) we have the following upper bound on the optimal threshold $\hat z^{(1)}_{th}$,
\be
\hat{z}^{(1)}_{th}<Z(C_{u0})=\frac{-\frac{Ae}{\sigma^2+\sigma_0^2}+\sqrt{\frac{A^2e^2}{(\sigma^2+\sigma_0^2)^2}-\left(\frac{1}{\sigma^2+\sigma_0^2}-\frac{1}{\sigma_0^2}\right)\left(\frac{A^2e^2}{\sigma^2+\sigma_0^2}-2\log\frac{C_{u0}\sigma_0}{\sqrt{\sigma^2+\sigma_0^2}}\right)}}{\frac{1}{\sigma_0^2}-\frac{1}{\sigma^2+\sigma_0^2}}.
\ee

On the other hand, the lower bound on the optimal threshold $\hat z^{(1)}_{th}$ can also be obtained based on Appendix B. Note that for $0 < z_{th} < Ae$, we have $\frac{1-p_0}{1-p_1} < \frac{1-\gamma_b}{1 - \gamma_t}$ and $\frac{p_0}{p_1} < 1$, and thus
\be\label{eq.threshold_lower}
\frac{(1-\gamma_t)G_0+\gamma_tG_1}{(1-\gamma_b)G_0+\gamma_bG_1}=\frac{\log\frac{1-p_0}{1-p_1}-\log\frac{p_0}{p_1}}{\frac{1-p_0}{1-p_1} -\frac{p_0}{p_1}}>\frac{\log\frac{1-\gamma_b}{1-\gamma_t}}{\frac{1-\gamma_b}{1-\gamma_t}-1}\dff C_{2},
\ee
which leads to $\hat z^{(1)}_{th}>Z(C_{l0})$, where $C_{l0}=\frac{\gamma_t-C_{2}\gamma_b}{C_{2}(1-\gamma_b)+\gamma_t-1}$.

We perform the same procedure for $D(p_1||p_0)$ and obtain another constant pair $(C_{l1},C_{u1})$, such that for the optimal detection threshold that maximizes $D(p_1||p_0)$ (denoted as  $\hat z^{(2)}_{th}$) we have that $Z(C_{l1}) < \hat z^{(2)}_{th} < Z(C_{u1})$. Let
\be
C_l&\dff&\max\{C_{l0},C_{l1}\}; \nonumber \\
C_u&\dff&\min\{C_{u0},C_{u1}\}.
\ee
We have that the minimum KL distance $\min\{D(p_0||p_1),D(p_1||p_0)\}$ is strictly increasing with respect to $z_{th}$ for $z_{th}\in[0,Z(C_l))$ and strictly decreasing with respect to $z_{th}$ for $z_{th}\in(Z(C_u),Ae]$, which reveals that the optimal threshold $\hat z_{th} = \arg \max_{z_{th}}\min\Big\{D\left(p_0||p_1\right), D\left(p_1||p_0\right)\Big\}$ cannot locate in the range of $[0,Z\left(C_l\right))\bigcup(Z\left(C_u\right),Ae]$. Therefore we have
\be
Z(C_l)\le \hat{z}_{th}\le Z(C_u).
\ee
\subsection{Proof of $(\ref{eq.threshold_upper})$ and (\ref{eq.threshold_lower})}
Consider the function
\be
G(a,b)=\frac{\log b-\log a}{b-a} \ \ \ \ \mbox{for}  \ \  b>a.
\ee
We have the following on the partial derivatives with respect to $a$ and $b$,
\be
\frac{\partial G}{\partial a}&=&\frac{1-\frac{b}{a}+\log\frac{b}{a}}{(a-b)^2}<0,\nonumber \\
\frac{\partial G}{\partial b}&=&\frac{1-\frac{a}{b}+\log\frac{a}{b}}{(a-b)^2}<0.
\ee
which shows that $G(a,b)$ is strictly decreasing with respect to $a$ and $b$.

\bibliographystyle{./IEEEtran}
\bibliography{./mybib}
\end{document}